\documentclass[aps,prl,twocolumn]{revtex4}
\usepackage{graphicx}

\def\vec#1{{\bf #1}}

\begin{document}

\title{Crossover from Conserving to Lossy Transport in Circular Random Matrix
Ensembles}

\author{Steven H. Simon${}^a$ and Aris L.  Moustakas${}^b$}

\affiliation{${}^a$Lucent Technologies, Bell Labs, Murray Hill, NJ, 07974, USA \\
${}^b$Department of Physics, University of Athens,
Panepistimiopolis, Athens 15784, Greece}

\begin{abstract}
In a quantum dot with three leads the transmission matrix $\vec
t_{12}$ between two of these leads is a truncation of a unitary
scattering matrix $\vec S$, which we treat as random.   As the
number of channels in the third lead is increased, the constraints
from the symmetry of $\vec S$ become less stringent and $\vec
t_{12}$ becomes closer to a matrix of complex Gaussian random
numbers with no constraints.   We consider the distribution of the
singular values of $\vec t_{12}$, which is related to a number of
physical quantities.
%When time reversal is broken this problem
%gives a realization of the Jacobi Unitary Ensemble.
\end{abstract}

\maketitle

Random matrix theory has enjoyed enormous success in a wide range
of fields of physics\cite{GuhrReview}.  One impressive success has
been in mesoscopic physics\cite{BeenakkerReview}, where the
scattering matrix between two leads (a source and a sink) of a
disordered quantum dot has been modelled as a random matrix. Such
scattering matrices can be from the circular unitary ensemble (CUE
or unitary matrices) if time reversal symmetry is broken, the
circular orthogonal ensemble (COE or unitary symmetric matrices)
if time reversal is unbroken, or the circular symplectic ensemble
(CSE or unitary self dual matrices) if spin-orbit scattering is
present and time reversal is unbroken\cite{BeenakkerReview}.
Another success of random matrix theory has been in the arena of
radio communications\cite{Tulino,MoustakasScience} where
transmission matrices between transmitting and receiving antennas
have been modelled as random matrices  with all matrix elements
being independent and identically distributed ($i.i.d.$) complex
Gaussian. These two examples are, in fact, quite closely related
to each other in that both stem from wave transmission problems.
The fundamental difference between these two cases is that the
former is ``conserving" while the latter is ``lossy". When an
electron enters a dot, it must exit one of the two leads whereas
in the radio communication problem most of the photons which are
transmitted are lost to the environment and neither reach the
receiver nor return to the transmitter.   In this manuscript we
will study the crossover between these two cases by studying a
quantum dot with a ``third lead" grounded at the same voltage as
the sink of the dot. As the number of channels to ground through
the third lead is increased (i.e., the amount of loss is
increased), the statistics of transmission between the source and
the sink become more and more similar to the $i.i.d.$ complex
Gaussian case {\it independent} of the underlying symmetry (COE,
CUE, or CSE). We note that this application of a ``third lead" is
quite distinct from that used previously in the mesoscopic
literature to model dephasing\cite{BeenakkerReview}. In that case,
a constraint is imposed that no current may flow out the third
lead; whereas in the present case, we allow current to flow out
the third lead, giving us true loss of electrons. We will not
consider dephasing here, so this work strictly applies to dots at
zero temperature.

We thus consider a quantum dot with three leads, a source (lead 1)
at voltage $V$, a grounded sink (lead 2) at voltage 0, and a loss
lead (lead 3) also grounded at voltage 0. We assume there are
$N_i$ fully open channels going in (and out of) lead $i$. It is
useful to define
\begin{eqnarray}
&N=N_1 + N_2 + N_3&~,  \\
    N_m = &\min(N_1,N_2)  ~~~~~, ~~~~~  N_{\Delta}& = |N_1 - N_2|.
\end{eqnarray}
The $N \times N$ scattering matrix between all three leads can be
written in block form as
\begin{equation}
    {\bf S} = \left(\begin{array}{c|c|c} {\bf r}_{11} & {\bf t}_{12} & {\bf t}_{13} \\ \hline
                              {\bf  t}_{21} & {\bf r}_{22}  & {\bf t}_{23} \\ \hline
                               {\bf t}_{31} & {\bf t}_{32} & {\bf r}_{33} \end{array} \right)
\end{equation}
where ${\bf r}_{ii}$ is $N_i \times N_i$ and ${\bf t}_{ij}$ is
$N_i \times N_j$. A quantity of interest is the $N_1 \times N_2$
matrix $\vec t_{12}$.   Making a singular value decomposition
(SVD) we have $\vec t_{12} = \vec U \vec D \vec V$ where $\vec U$
is an $N_1$ dimensional unitary matrix, $\vec V$ is an $N_2$
dimensional unitary matrix, and $\vec D$ is an $N_1 \times N_2$
matrix with the (real) singular values $d_i > 0$ along the $N_m$
diagonal elements and zeros elsewhere.   More physically relevant
perhaps is the Hermitian matrix
\begin{equation}
\label{eq:Tdef} {\bf T}_{12} = {\bf t}_{12}^\dagger
 {\bf t}_{12} = \vec V^\dagger \vec D^\dagger \vec D \vec V
\end{equation}
whose $N_m$ nonzero eigenvalues $z_i = d_i^2 $ correspond to the
squares of the singular values of $\vec t_{12}$.  It is convenient
to work in terms of the eigenvalues $\{ z_i \}$ rather than the
singular values $\{ d_i \}$. Many physical quantities are given in
terms of these eigenvalues\cite{BeenakkerReview,Langen}.  For
example, the current out of lead $2$ is given by $I_2 = g_{12} G_0
V$ where $G_0 = 2 e^2/h$ and
\begin{equation}
g_{12} =  {\rm Tr}[ {\bf T}_{12}] = \mbox{$\sum_{i=1}^{N_m} \,
z_i$}
\end{equation}
Similarly the shot noise of the current at lead 2 can be
shown\cite{BeenakkerReview,Langen} to be given by $2 e V G_0 s$
with $s={\rm Tr}[ {\bf T}_{12} ({\bf 1} -{\bf T}_{12})] =
\sum_{i=1}^{N_m} z_i (1 - z_i)$. These physical quantities are
examples of so-called ``linear statistics"\cite{BeenakkerReview}
whose expectations can be written as an integral over the
eigenvalue density $\rho(z)$ of the nonzero eigenvalues of $\vec
T_{12}$. Other linear statistics can also be shown to give
physically interesting quantities related to higher moments of the
current\cite{BeenakkerReview,Langen,Levitov}.

We note that a typical element of the matrix ${\bf S}$ has
magnitude $1/N^{1/2}$ (since ${\bf SS}^\dagger = {\bf 1}$), so
increasing $N_3$ has the effect of decreasing all the elements of
${\bf t}_{12}$. In addition to this normalization change,
qualitatively we should expect that the symmetry (COE, CUE, or
CSE) constraints on $\bf S$ become less visible in the submatrix
${\bf t}_{12}$ as $N_3$ is increased, since the constraints are
more easily satisfied by the large number of additional variables
in the other blocks of $\bf S$. Thus, we expect that ${\bf
t}_{12}$ should start to look like an $i.i.d.$ random complex
Gaussian matrix as $N_3$ becomes large.  When ${\bf t}_{12}$ is
$i.i.d.$ random complex Gaussian, the corresponding ${\bf T}_{12}$
is known as a ``complex Wishart"
matrix\cite{Wishart,Edelman,Forrester}.

As is frequently the case in random matrix problems, it is easiest
to make further progress if one focuses on $\bf S$ being from the
unitary ensemble (broken time reversal invariance).  In this case
we use a method analogous to that of Ref.~\cite{Sommers} to
determine the joint probability density (JPD) of the nonzero
eigenvalues $\{ z_i \}$ of $\vec T_{12}$.    We start by writing
the probability density ${\cal P}(\vec S) \propto \delta(\vec
S^\dagger \vec S - \vec 1)$ where we have an $N^2$ dimensional
complex $\delta$ function to enforce unitarity. Analogous to
\cite{Sommers} we integrate out $\vec r_{33}, \vec r_{22}, \vec
t_{31}, \vec t_{32}, \vec t_{21}, \vec t_{23}$ to obtain
\begin{equation}
   {\cal  P}(\vec t_{12}) \propto \!\! \int \!\! d\vec t_{13} \!\!\int \!\! d\vec
     r_{11}\,
     \delta(\vec r_{11}^\dagger \vec r_{11} + \vec t_{12}^\dagger \vec t_{12} + \vec t_{13}^\dagger \vec t_{13} - \vec 1)
 \end{equation}
with the delta function being $N_1^2$ complex dimensional, and the
integrals being $N_1 (N_1 + N_3)$ complex dimensional.  Writing
$\vec t_{12}^\dagger \vec t_{12}$ in terms of its SVD as in
Eq.~\ref{eq:Tdef}, the Jacobian of this change of variables is
given by $ J \sim \Delta(\{ z \} )^2\, \prod_i z_i^{N_{\Delta}} $
where $\Delta(\{ z \})$ is the Vandermonde determinant
\begin{equation}
\label{eq:vander}
    \Delta(\{ z \}) = \prod_{1 \leq i < j \leq N_m} (z_i - z_j) \,\,\, = \,\,\, \det[
    z_i^{j-1}].
\end{equation}
We can then integrate out $\vec U$ and $\vec V$ to obtain
\begin{equation}
    {\cal P}(\{z \}) \propto  J \!\! \int \!\! d\vec t_{13} \!\!\int \!\! d\vec
     r_{11}\,
     \delta(\vec r_{11}^\dagger \vec r_{11} + \vec t_{13}^\dagger \vec t_{12}  + \vec z  - \vec 1)
\end{equation}
where $\vec z$ is a diagonal matrix of the $N_m$ nonzero
eigenvalues $z_i$ (with zeros along the remaining diagonals if
$N_1 > N_2$).  Performing the remaining integrals yields the final
result for the JPD
\begin{equation}
{\cal P}(\{z \}) \propto \Delta(\{ z \})^2  \prod_{i=1}^{N_m}
 w(z_i)  \label{eq:JPD}
 \end{equation}
 with weight function
 \begin{equation}
 \label{eq:weight}
    w(z) = z^{N_{\Delta}} (1 - z)^{N_3}\,  \Theta(z) \Theta(1 -z)
\end{equation}
where the step function $\Theta(x)$ is 1 for  $x > 0$ and is zero
otherwise. In the limit of large $N_3$ (with $N_\Delta$ and $N_m$
fixed) the factor $(1 - z)^{N_3}$ becomes $e^{-z N_3}$  ($\sim
e^{-z N}$ since $z (N_1 + N_2)$ becomes small) as expected for the
limiting complex Wishart
distribution\cite{Wishart,GuhrReview,Forrester,Edelman}.

It is interesting that Eqs. \ref{eq:JPD}, \ref{eq:weight} are
similar in form to the distribution found for the (complex) {\it
eigenvalues} of a (square) truncated unitary matrix found by
Ref.~\cite{Sommers}.   Here, in contrast, we are looking at the
{\it singular} values of the truncated unitary matrix (${\bf
t}_{12}$).   Despite the similarity of these cases, we will see
below that the distribution of the singular values  of ${\bf
t}_{12}$ is {\it very} different from that of the absolute value
of its eigenvalues.

The Vandermonde determinant can be recast as a Slater determinant
of a set of wavefunctions $\{ \phi_i \}$ which are orthonormal
with respect to the weight function $w(z)$
\begin{equation}
    \phi_i(z) = C_i \, P_i^{(N_3, N_\Delta)}(2 z - 1)
\end{equation}
for $i=0 \ldots N_m-1$ where $P_i^{(a,b)}(x)$ is the Jacobi
polynomial\cite{Abromowitz}, and
$
    C_i^2 = [(2 i + N_3 + N_\Delta+1)
    i! (i + N_3 + N_\Delta)! ] / [(i+N_3)!  \, \, (i+N_\Delta)!]
$.  Thus our distribution is a realization of the so-called Jacobi
Unitary Ensemble (JUE)\cite{Edelman,Forrester}. Interestingly this
ensemble occurs in a completely different random matrix problem
when one considers the eigenvalues of $A (A+B)^{-1}$ where $A$ and
$B$ are both complex Wishart matrices\cite{Edelman,Forrester}.  In
the case of the quantum dot with two leads ($N_3 = 0$) it has been
previously pointed out\cite{Forrester} that the eigenvalue
distribution is a special case of the JUE. Here we have shown that
all possible JUEs may be realized in three lead dots.

Using standard techniques\cite{Forrester,GuhrReview}, once we have
our JPD in this orthogonal polynomial form, we can write the
kernel $K(z_1, z_2)$ as
%\begin{eqnarray}
% & &   K(z_1,z_2)/ \sqrt{w(z_1) w(z_2)} = \sum_{i=0}^{N_m-1} \phi_i(z_1)
%    \phi_i(z_2) ~~~~~~~ \\
% & &        =    \left[ \frac{\phi_{N}(z_1) \phi_{N-1}
%    (z_2) - \phi_{N-1}(z_1) \phi_N(z_2)}{q ( z_1 - z_2)} \right]
%    \label{eq:resummed}
%\end{eqnarray}
%with \begin{equation} q = \frac{N \sqrt{N^2 - 1}}{\sqrt{N_m
%(N-N_m) (N_m + N_3) (N_m + N_\Delta)}}
%\end{equation}
%where the resummation here is performed by the Christoffel-Darboux
%formula.
\begin{equation}
  K(z_1,z_2) =  \sqrt{w(z_1) w(z_2)}  \sum_{i=0}^{N_m-1} \phi_i(z_1)
    \phi_i(z_2)
\end{equation}
and this series can be resummed using the Christoffel-Darboux
formula as usual\cite{Forrester,GuhrReview} if desired.   All
correlation functions between the eigenvalues $\{ z_i \}$ can be
expressed analytically in terms of this
kernel\cite{Forrester,GuhrReview}, the simplest example of which
is the normalized single eigenvalue density $\rho(z) = K(z,z)/N_m$
which is shown in Fig.~1.  For comparison, the orthogonal case is
also shown (data from numerical monte-carlo simulations).  In both
cases, the distribution clearly converges to the complex Wishart
form\cite{Wishart,Edelman,Forrester}for large $N_3$.  Similar
convergence to the Wishart form is obtained from numerical
simulations for the symplectic case (not shown).

\begin{figure}[htbp]
\hspace*{-20pt} \vspace*{15pt}
\scalebox{.38}{\includegraphics*{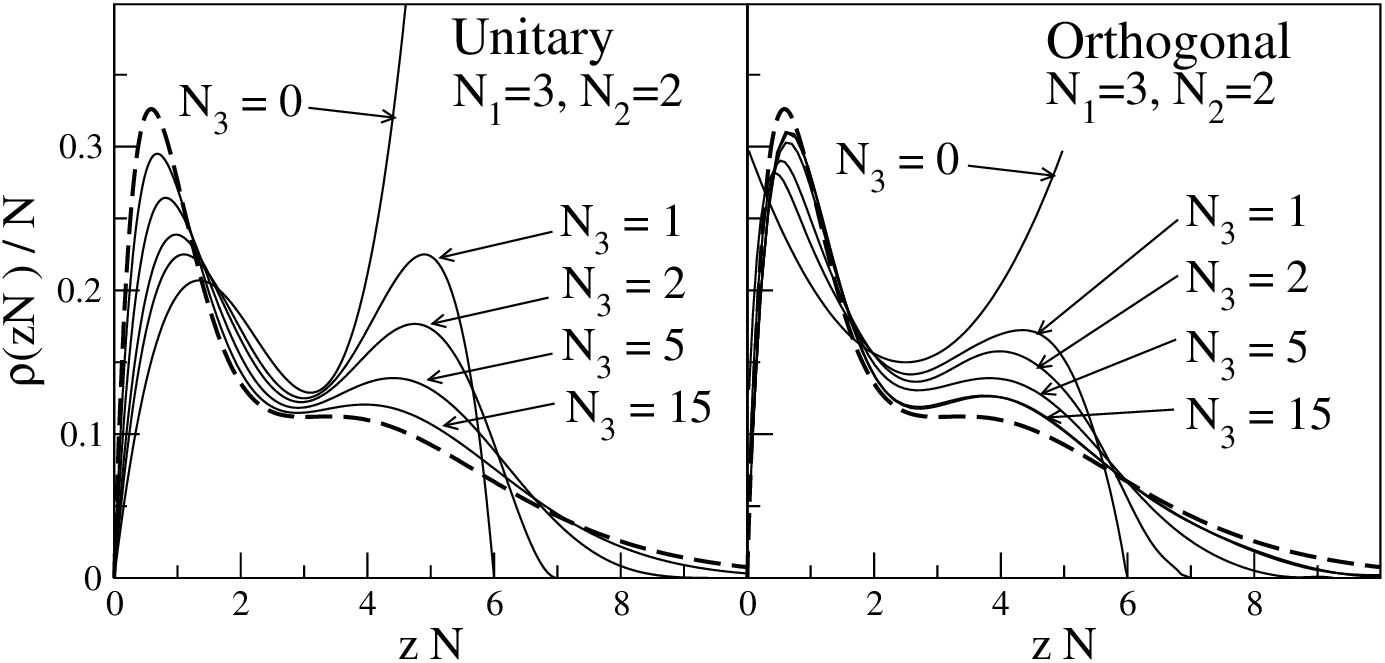}} \hspace*{-20pt}
\scalebox{.38}{\includegraphics*{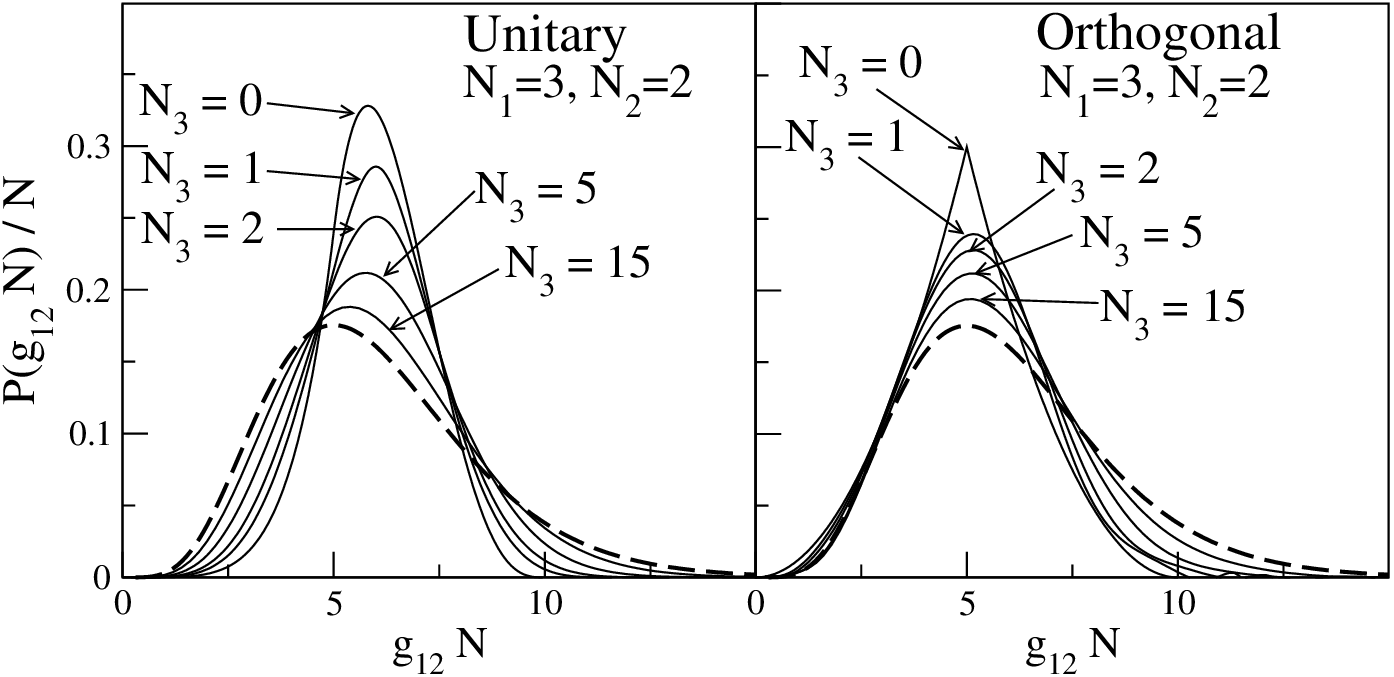}} \caption{{\bf Top:}
Scaled eigenvalue density $\rho(z N)/N$ of $\vec T_{12} = \vec
t_{12}^\dagger \vec t_{12}$. {\bf Bottom:} Scaled distribution
$P(g_{12} N)/N$ with $g_{12}={\rm Tr}[\vec T_{12}]$. In all cases
data is shown for $N_1 = 3, N_2=2$ and $N_3=0,1,2,5,15$ (note
$N=N_1 + N_2 + N_3$). The unitary cases are analytic as discussed
in the text. The orthogonal cases ($N_3 \neq 0$) are numerical and
the curves have been smoothed for clarity. The orthogonal $N_3=0$
case is also analytic as given by
Refs.~\cite{BeenakkerReview,Baranger}. The dashed curves are the
Wishart limit\cite{Wishart} corresponding to $N_3 \rightarrow
\infty$ and is the same for both unitary and orthogonal cases. The
eigenvalue density is only nonzero for $z < N$, and $P(g_{12} N)$
is only nonzero for $g_{12}  < N_m = 2$.}
\end{figure}

We can also ask about the distribution of $g_{12}={\rm Tr}[\vec
T_{12}]$ which, as mentioned above, corresponds to the conductance
of our dot.  We have
\begin{eqnarray}
P(g_{12}) &=&   \int d\{z_i \} \,\, \delta\left(g_{12} -
\mbox{$\sum_i$} z_i \right) \, {\cal P}(\{ z \})
\\
 &=&  \!\! \int \! \frac{dk}{2 \pi} \!\!  \int \! d\{z_i \} \, e^{-i k
\left( g_{12} - \sum_i z_i \right)} \, {\cal P}(\{ z \})
\end{eqnarray}
It is now convenient to rewrite the $\Delta(\{ z_i \})$ in $P(\{ z
\})$ in its determinant form (Eq.~\ref{eq:vander}) and expand this
determinant as an antisymmetrized product.  We can then perform
the integrals to obtain
\begin{equation}
\label{eq:detexp}
    P(g_{12}) = \frac{1}{\det[\vec F(0)]}  \int \frac{dk}{2 \pi} e^{-i k \, g_{12}} \det[ \vec F(k) ]
\end{equation}
where $\bf F$ is an $N_m \times N_m$ matrix with elements
\begin{eqnarray}
   F_{ab}(k)
            &=& \int_0^1 \, dz \, z^{a + b -2 + N_{\Delta}} \, (1 - z)^{N_3} \,  e^{i
            k z} \label{eq:complicatedint}
\end{eqnarray}
The integrals of Eq.~\ref{eq:complicatedint} although tedious can
always be done, resulting in the form
\begin{equation}
    F_{ab}(k) = Q_{ab}(1/k)  +  e^{i k}  R_{ab}(1/k)
\end{equation}
where $Q_{ij}$ and $R_{ij}$ are simple polynomials of degree
$a+b-1+N_\Delta + N_3$, and $F_{ij}(0)$ remains finite.  We then
have
\begin{equation}
    \det[\vec  F(k)] =  \sum_{n=0}^{N_m}  D_n(1/k)  e^{i k n}
\end{equation}
where the $D_n$'s are polynomials of order $N_m (N - N_m)$.  The
Fourier transform in Eq.~\ref{eq:detexp} can easily be carried out
analytically resulting in $P(g)$ being of the form
\begin{equation}
    P(g_{12})= \sum_{n=0}^{N_m} L_n(g_{12})  \Theta(g_{12}-n)
\end{equation}
where each $L_n$ is an $N_m (N - N_m)-1$ degree polynomial.  The
detailed forms of these polynomials can easily be evaluated on a
case-by-case basis (trivially with Mathematica).  In the limit of
large $N_3$,  the $(1-z)^{N_3}$ factor in the integral in
Eq.~\ref{eq:complicatedint} can be replaced by $e^{-z N_3}$ and
the upper limit of the integral can be extended to $\infty$.   We
then have $F_{ab}(k) = (N_\Delta + a + b - 2)! (N_3  - ik)^{(1 - a
-b - N_\Delta)}$.  The determinant is then $\det \vec F(k) \sim
(N_3 - ik)^{-N_1 N_2}$ resulting in $P(g_{12}) \sim g_{12}^{N_1
N_2 -1} e^{-g_{12} N_3}
 \approx g_{12}^{N_1 N_2 -1} e^{-g_{12} N}$, which is known as the
Rayleigh distribution --- the characteristic distribution for the
trace of a complex Wishart matrix.

In Fig.~1, examples of the distribution $P(g_{12})$ are shown. The
results in the unitary case are analytic as described above,
whereas the orthogonal data has been generated numerically.
Convergence to the complex Wishart (Rayleigh) limit is again clear
in both cases. Similar convergence to this limiting form is
obtained from numerical simulations for the symplectic case (not
shown).

As with most random matrix problems, considerable simplification
occurs in the limit of large matrices.  We will thus consider the
limit of $N_1, N_2, N_3 \rightarrow \infty$ while keeping the
ratios $N_i/N $ fixed.
%In this case, the asymptotics of the
%Jacobi polynomials are given by \cite{JacobiLimit}.
We proceed by considering Eq.~\ref{eq:JPD} as a classical
one-dimensional gas with Boltzmann weight given by ${\cal
P}(\{z\})$. Thus, we have a one particle potential given by
$e^{-V(z)} = w(z)$, and two particle interaction $e^{-W(z_i,z_j)}
= (z_i - z_j)^2$.  Using standard
methods\cite{GuhrReview,Forrester,Eynard}, we derive the saddle
point equations
\begin{equation}
V'(z_i) =   \sum_{j \neq i} \frac{2}{z_i - z_j}
\end{equation}
We then introduce the Green's function
\begin{equation}
    G(z) = \frac{1}{N_m} \sum_i \frac{1}{z - z_i} = \int dz'
    \, \frac{\rho(z')}{z - z'}
\end{equation}
with $\rho(z) = \pi^{-1} \, {\rm Im}\, G(z+i 0^+)$ being the
normalized single eigenvalue density. It is easy to
verify\cite{GuhrReview,Forrester,Eynard} that to order $1/N$ we
have $G(z)^2 = \sum V'(z_i)/(z - z_i)$ which we can solve
analytically resulting in the eigenvalue density
\begin{equation}
\label{eq:largeNlimit} \rho(z) = \frac{N}{N_m}
\frac{\sqrt{(z - z_-) (z_+ - z)}}{ 2 \pi z (1 - z)}
\end{equation}
for $z_- < z < z_+$ and $\rho(z) = 0$ otherwise, where
\begin{equation}
    z_{\pm} = N^{-2} \left(\sqrt{N_1 (N - N_2)} \,\,\,  \pm \,\,\,  \sqrt{N_2 (N - N_1)}
   \,\, \right)^2
\end{equation}
We have also derived this simple form by using the
free-probability\cite{Free} method, which we will present in a
separate paper\cite{Us}.  In fact, we also show in that work that
this form is correct in the large $N$ limit {\it independent} of
whether the original scattering matrix $\vec S$ is from the COE,
CUE, or CSE. This result should not be surprising, as it is well
known\cite{GuhrReview,BeenakkerReview} that for many quantities of
physical interest  (such as the average conductance, etc), the
ensembles differ from each other only at order unity whereas
leading terms will be order $N$. These small differences are known
as the weak localization corrections.

In Fig.~2 we show how the eigenvalue distributions converge to the
large $N$ limit, holding $N_i/N$ fixed.  Analytic results are
shown for the unitary case, and numerical results are shown for
the orthogonal case.  In both cases we see convergence to the
limiting form of Eq.~\ref{eq:largeNlimit} as expected. Similar
convergence to this limit is seen numerically for the symplectic
case (not shown).

\begin{figure}[tbp]
\hspace*{-20pt}  \scalebox{.38}{\includegraphics*{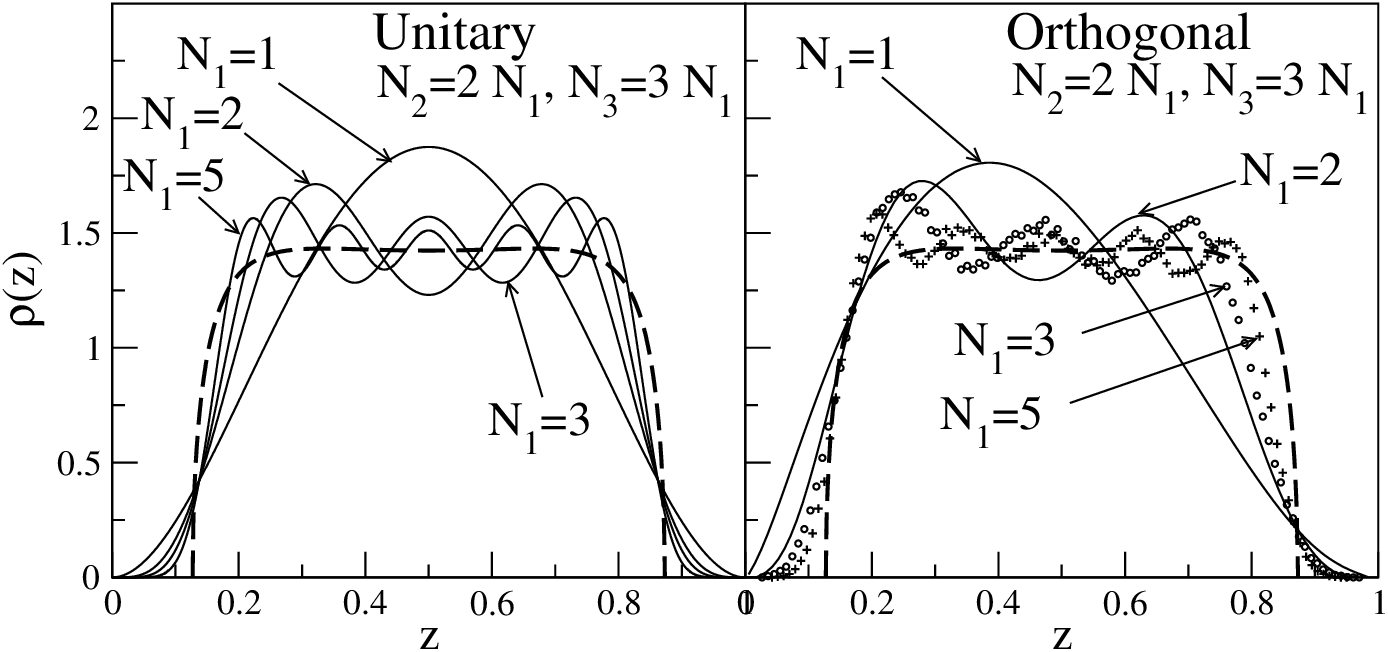}}
\caption{Eigenvalue density $\rho(z)$ of $\vec T_{12} = \vec
t_{12}^\dagger \vec t_{12}$ for $N_1/N=1/6$, $N_2/N=3/6$ and
$N_3/N=2/6$ for $N_1=1,2,3,5$ ($N=6,12,18,30$). The unitary case
is calculated analytically as described in the text. The
orthogonal case is calculated numerically.   For $N_1=1,2$ the
orthogonal case numerical data has been smoothed for clarity.  For
$N_1=3$ (circles) and $N_1=5$ (pluses), smoothing is found to
destroy the oscillations, so the raw histogrammed data is shown.
The dashed line is the large $N$ limit (Eq.~\ref{eq:largeNlimit})
to which both the unitary and orthogonal cases converge.}
\end{figure}

As  mentioned above, the problem of calculating the eigenvalue
distribution of $\vec T_{12}$, or equivalently the singular value
distribution of $\vec t_{12}$ appears similar to the problem of
calculating the complex eigenvalue distribution of a (square)
truncated random unitary matrix derived in \cite{Sommers}.   To
see the difference between these two problems, let us examine the
case of $N_1/N = N_2/N =n \leq 1/2$ in the large $N$ limit for
which both calculations are applicable. From our results, we find
the squared singular values density decreases monotonically to
zero at $z = z_+ = 4 n (1 - n)$.   In the work of \cite{Sommers},
however, it is found that the density of the norm squared of the
eigenvalue distribution is a  function that {\it increases}
monotonically, then drops to zero discontinuously at $z=n$.

One might attempt to guess at the general form of the JPD,
analogous to Eqs. \ref{eq:JPD} and \ref{eq:weight}, for the COE
and CSE.  Indeed, in the case of $N_3=0$ a simple generalization
exists\cite{BeenakkerReview,Baranger}  which involves changing the
exponent of $z$ in the weight $w(z)$ and changing the exponent of
the Vandermonde determinant in Eq.~\ref{eq:JPD} to $\beta=1$ for
the COE and to $\beta =4$ for the CSE case. However, for the $N_3
\neq 0$ problem no such simple form can apply, since for large
$N_3$ we always expect that $\vec T_{12}$ will be Wishart (ie,
$\vec t_{12}$ is simply an $i.i.d.$ complex Gaussian matrix) so
that there will always be quadratic ($\beta = 2$) eigenvalue
repulsion {\it independent} of whether $\vec S$ is from the COE,
CUE, or CSE. We thus leave the full solution of the JPD for the
COE and CSE as an open problem.

The authors are greatly indebted to A. Lamacraft and B. Simons for
pointing us to this crossover as an interesting problem to
examine.

\end{document}